\documentclass[twocolumn]{emulateapj}
\usepackage{psfig}
\usepackage{graphicx}

\advance \voffset by 0.00cm\relax

\shorttitle{IN-SPIRAL RATES OF DOUBLE NEUTRON STARS}
\shortauthors{KALOGERA ET AL.}

\begin{document}

\title{The Cosmic Coalescence Rates for Double Neutron Star Binaries}

\author{V.\ Kalogera\altaffilmark{1}, 
        C.\ Kim\altaffilmark{1}, D.\ R.\ Lorimer\altaffilmark{2},
        M.\ Burgay\altaffilmark{3}, N.\ D'Amico\altaffilmark{4,5}, 
        A.\ Possenti\altaffilmark{5,6}, R.\ N.\ Manchester\altaffilmark{7}, 
        A.\ G.\ Lyne\altaffilmark{2}, B.\ C.\ Joshi\altaffilmark{2,8}, 
        M.\ A.\ McLaughlin\altaffilmark{2}, M.\ Kramer\altaffilmark{2}, 
        J.\ M.\ Sarkissian\altaffilmark{7}, and F.\ Camilo\altaffilmark{9}
}

\affil{ $^{1}$ Northwestern University, Dept.\ of Physics \& Astronomy,
       2145 Sheridan Rd., Evanston, IL 60208\\ 
       $^{2}$ University of Manchester, Jodrell Bank Observatory, Macclesfield,
Cheshire SK11 9DL, UK\\
       $^{3}$ Universit{\'a} degli Studi di Bologna, Dipartimento di Astronomia, via Ranzani 1, 40127, Bologna, Italy\\
       $^{4}$ Universit{\'a} degli Studi di Cagliari, Dipartimento di Fisica, SP Monserrato-Sestu km 0.7, 09042 Monserrato, Italy\\
       $^{5}$ INAF - Osservatorio Astronomico di Cagliari, Loc. Poggio dei Pini, Strada 54, 09012 Capoterra, Italy\\
       $^{6}$ INAF - Osservatorio Astronomico di Bologna, via Ranzani 1, 40127, Bologna, Italy\\
       $^{7}$ Australia Telescope National Facility, CSIRO, PO Box 76, Epping, NSW 2121, Australia\\
       $^{8}$ National Center for Radio Astrophysics, PO Bag 3, Ganeshkhind, Pune 411007, India\\
       $^{9}$ Columbia Astrophysics Laboratory, Columbia University, 550 West 120$^{\rm th}$ Street, New York, NY 10027\\
       vicky, c-kim1@northwestern.edu; drl@jb.man.ac.uk; 
        burgay@tucanae.bo.astro.it;
        damico@ca.astro.it;   
        possenti@ca.astro.it;
        rmanches@atnf.csiro.au; 
        agl@jb.man.ac.uk; 
        bcj@ncra.tifr.res.in; mclaughl@jb.man.ac.uk;
        mkramer@jb.man.ac.uk;
        John.Sarkissian@csiro.au;fernando@astro.columbia.edu
}

\begin{abstract} 
This manuscript is an updated version of Kalogera et al. (2004) published in ApJ Letters 
to correct our calculation of the Galactic DNS coalescence rate. 
The details of the original erratum submitted to ApJ Letters are given in page 6 of this manuscript.

We report on the newly increased event rates due to the recent discovery of the highly relativistic binary pulsar J0737--3039 (Burgay et al.\ 2003). Using a rigorous statistical method, we present the calculations reported by Burgay et al.\,, which produce a coalescence rate for Galactic double neutron star (DNS) systems that is higher by a factor of $5-7$ compared to estimates made prior to the new discovery. Our method takes into account known pulsar-survey selection effects and biases due to small-number statistics. This rate increase has dramatic implications for gravitational wave detectors. For the initial Laser Interferometer Gravitational-Wave Observatory (LIGO) detectors, the most probable detection rates for DNS in-spirals are one event per $10-630$ yr; at 95\% confidence, we obtain rates up to one per 3 yr. For the advanced LIGO detectors, the most probable rates are $10-500$ events per year. These predictions, for the first time, bring the expectations for DNS detections by the initial LIGO detectors to the astrophysically relevant regime. We also use our models to predict that the large-scale Parkes Multibeam pulsar survey with acceleration searches could detect an average of four binary pulsars similar to those known at present. 
\end{abstract}

\keywords{binaries: close -- stars: neutron -- gravitational waves --
methods: statistical}

\section{INTRODUCTION}
\label{sec:intro}

For almost 30 years, close double neutron star (DNS) binary systems have been known to exist in the Galaxy as a small subset of the observed radio pulsar population (Hulse \& Taylor 1975; Wolszczan 1991). These systems lose orbital energy due to the emission of gravitational waves (Taylor \& Weisberg 1989, 2003; Stairs et al.\ 1998); the associated orbital in-spiral continues until the binary system coalesces, resulting in a burst of gravitational waves. DNS in-spirals are prime targets for gravitational-wave detection by the ground-based interferometers Laser Interferometer Gravitational-Wave Observatory (LIGO; Abramovici et al.\ 1992), GEO (Danzmann et al.\ 1995), and VIRGO (Caron et al.\ 1997). Event rate estimates are very important for the development of gravitational-wave interferometers (Thorne \& Cutler 2002).  They are based on estimates of Galactic rates and their extrapolation throughout a survey volume (Finn 2001), given the source strength and instrument sensitivity. For DNS binaries, Galactic rate estimates have been obtained using two very different methods. One is purely theoretical and involves models of binary evolution calibrated usually to the observationally determined supernova rate for the Galaxy. The other, more empirical, approach is based on the physical properties of the close DNS binaries known in the Galactic field and modeling of radio pulsar survey selection effects. For a review and details of both these approaches, see Kalogera et al.\ (2001, hereafter KNST) and references therein. The empirical method has generally provided us with better constraints on the coalescence rate (KNST), although the uncertainty still exceeds two orders of magnitude. This is primarily due to (1) the very small number (only two until recently) of close DNSs known in the Galactic field with merger times shorter than a Hubble time and (2) the implicit assumption that this small sample is a good representation of the total Galactic population (KNST).

Two recent developments make it appropriate to revisit the DNS merger
rate calculations. First, the discovery of the 2.4 hr DNS binary
PSR J0737--3039 in a large-area survey using the Parkes radio
telescope (Burgay et al.\ 2003) brings the number of known DNS systems to merge in the
Galactic field to three. With an orbital period of only 2.4 hr,
J0737--3039 will coalesce in only 85 Myr, a factor of 3.5 shorter than
the merger time of PSR B1913+16. This
immediately hints towards a possible significant increase of the
coalescence rate (Burgay et al.\ 2003). Second, a novel statistical method
has been developed by Kim, Kalogera \& Lorimer (2003, hereafter
KKL) that automatically takes into account statistical biases inherent
in small-number samples, like the relativistic DNS binaries, and 
in addition allows us to quantify our expectation that the actual DNS binary coalescence rate has a particular value, given the current observations.

In this Letter, we use this statistical method and investigate in
detail the effect of this new DNS discovery on the estimates of
Galactic DNS in-spiral rates and its implications for
gravitational wave detection in this decade. We summarize 
the method in \S\,\ref{sec:method} and discuss the resulting
Galactic in-spiral rate in \S\,\ref{sec:results}. In
\S\,\ref{sec:pmb} we use our models to make predictions for the
expected number of DNS binaries that the Parkes Multibeam
(PMB) survey (e.g. Manchester et al.\ 2001) could detect when acceleration
searches are completed, and in \S\,\ref{sec:discussion} we discuss the implications of our results for the detection rates of upcoming gravitational-wave detectors.

\section{METHOD FOR RATE CALCULATION}
 \label{sec:method}

Until recently, estimates of DNS coalescence rates provided a range of
possible values without any information on the likelihood of these
values. KKL presented a newly developed statistical analysis that
allows the calculation of a {\em probability distribution} for  rate estimates and the determination of confidence intervals associated with the rate estimates. The method can be applied to any radio pulsar population (see KKL for the two close DNS known at the time and Kim et al.\ 2004 for close binaries with a pulsar and a massive white dwarf companion). Here we update the results of KKL, taking into account the recent discovery of the new DNS binary PSR J0737--3039 (Burgay et al.\ 2003).

The method is described in detail in KKL, but we briefly summarize the
main elements here. The method involves the simulation of selection
effects inherent in all relevant radio pulsar surveys and a Bayesian
statistical analysis for the probability distribution of the in-spiral
rate estimates. The small-number bias and the effect of the faint end of the
pulsar luminosity function, previously identified as the main sources
of uncertainty in rate estimates (KNST), are {\it implicitly included}
in this analysis.

For a model Galactic pulsar population with an assumed spatial and
luminosity distribution, we determine the fraction of the total
population which are actually {\it detectable} by current large-scale
pulsar surveys. In order to do this, we calculate the effective
signal-to-noise ratio for each model pulsar in each survey, and
compare this with the corresponding detection threshold. Only those
pulsars which are nominally above the threshold count as
detectable. After performing this process on the entire model pulsar
population of size $N_{\rm tot}$, we are left with a sample of $N_{\rm
obs}$ pulsars that are nominally detectable by the surveys. By
repeating this process many times, we can determine the probability
distribution of $N_{\rm obs}$, which we then use to constrain the
population and with a Bayesian analysis derive the probability
 expectation that the actual Galactic DNS in-spiral rate takes on a particular value, given the observations. More details are
given in \S\,2 of Lorimer et al.\ (1993) and in KKL.

When this method was first developed (KKL), it was shown that, although
the shape of the probability distribution of rate estimates is very robust, the
rate value at peak probability systematically depends primarily on the
characteristics of the radio pulsar luminosity function: its slope and
the physical minimum luminosity of pulsars. Both of these are
constrained by the general pulsar population (see Cordes
\& Chernoff 1997), but we explore the dependence of our results on the
assumed values.

Here we consider the same set of pulsar population models as in KKL,
but we choose model 6 as our reference model in view of the recent
discovery of very faint pulsars (for a review, see Camilo 2003).
With the addition of the new DNS binary PSR J0737--3039, our
calculations differ from those in KKL in two main ways: (1) the
latest Parkes survey that led to the discovery of the new system
(Burgay et al.\ 2003) is included, and (2) we calculate and account for
the effects of Doppler smearing for DNS binaries akin to J0737--3039
by creating fake time series for a variety of orbital phases (see KKL
for details).  Even for a $\simeq4.5-$min integration (Burgay et al.\ 2003), this effect alone reduces the average signal-to-noise ratio of a 2.4-hr DNS binary by 35\%.

The statistical analysis presented in KKL has been extended to account for three systems (see Kim et al.\ 2004). In our calculations we adopt a total lifetime for J0737--3039, defined as the sum of the current age and the remaining lifetime until the final coalescence, equal to $100 + 85 = 185$ Myr (Burgay et al.\ 2003). In the absence of detailed beam observations for the new binary, we adopt a beaming factor of $\simeq 6$ equal to the average of the two observationally constrained beams of the previously known DNS binaries (KNST; see also Burgay et al.\ 2003). We do note, however, that studies of other known recycled pulsars (the majority of them strongly recycled, spinning faster compared to the DNSs by about an order of magnitude) have shown that beaming fractions can vary significantly (Kramer et al.\ 1998). It is important to keep in mind that any uncertainties in the beaming factor proportionally affect the rate estimates, but not the rate increase factors derived here.

\section{GALACTIC IN-SPIRAL RATE}
 \label{sec:results} 

\begin{figure}
\begin{center}
\psfig{figure=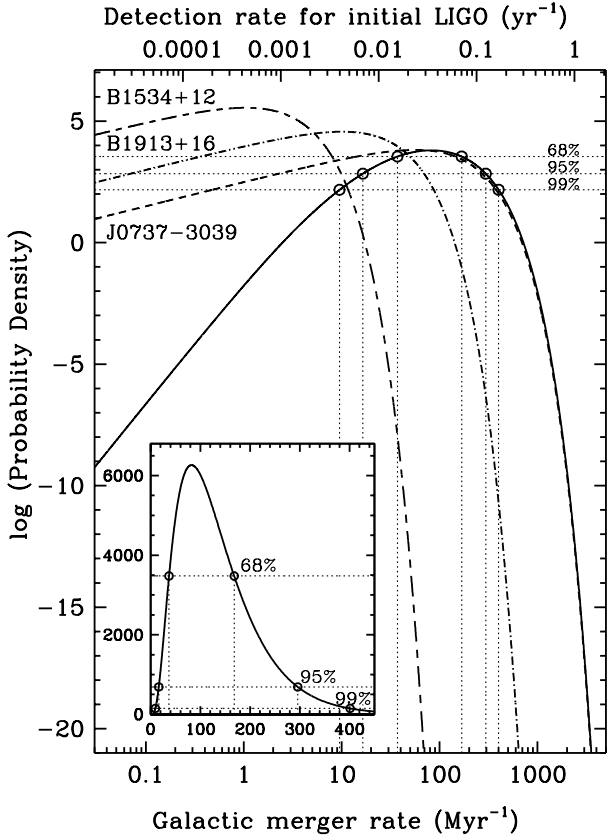,width=3.3in}
\caption{Probability density function that represents our expectation that the actual DNS binary
merger rate in the Galaxy ({\it bottom axis}) and the predicted initial LIGO
rate ({\it top axis}) take on particular values, given the observations. The curves shown are calculated assuming our reference model parameters (see text).
The solid line shows the total probability density
along with those obtained for each of the three binary systems ({\it dashed lines}). {\it Inset}: Total probability density, and corresponding 68\%, 95\%, and 99\% confidence limits, shown in a linear scale.
}
\label{fig:pdf}
\end{center}
\end{figure}

For our reference pulsar model (with a radio luminosity function
consistent with current pulsar observations; Cordes \& Chernoff 1997;
Camilo 2003), we find the most likely value of the total coalescence
rate to be ${\cal R} = 83$ Myr$^{-1}$. The ranges of values at 68\%
and 95\% confidence intervals are 40--170 and 20--290\,Myr$^{-1}$, respectively. The width of these ranges are somewhat smaller than previous estimates (the ratio between upper and lower limits at 68\% confidence interval is 4.4 cf.~5.7 found by KKL), confirming the expectation that a bigger observed sample would reduce the uncertainty
in the rate estimates (KNST). The new value for ${\cal R}$ is a factor
of 6.4 higher than found by KKL. From the resulting probability
distribution shown in Figure \ref{fig:pdf}, it is clear that
J0737--3039 dominates the total rate over the other two systems. This
is due to two separate factors: (1) The estimated total number of DNS
binaries similar to J0737-3039 (1700) is far higher than those of each
of the other two systems (700 for B1913+16 and 500 for B1534+12). This
is mainly due to the shorter pulsar spin and binary orbital period of J0737--3039, which results in a significant Doppler smearing and efficiently ``hides'' them in the Galaxy. (2) The total lifetime of J0737--3039
(185 Myr) is significantly shorter than those of the other two (365 Myr for B1913+16 and 2.9 Gyr for B1534+12).

%%%%%%%%%%%%%%%%%%%%%%%%%%%%%%%%%%%%%%%%%%%%%%%%%%%%%%%%%%%%%%%%%%%%%%%
%% Table 1. (table1)

 \begin{deluxetable}{crcrr}
% \begin{deluxetable}{llcll}
 \label{tab:results}
 \tablewidth{1000pc}
 \tabletypesize{\footnotesize}
 \tablecaption{Estimates for Galactic in-spiral rates
and predicted LIGO detection rates (at 95\% confidence)
for different population models}
 \tablehead{
\colhead{Model$^{a}$} &
\colhead{${\cal R}$} &
\colhead{IRF$^{b}$}&
\multicolumn{2}{c}{${\cal R}_{\rm det}$ of LIGO} \\
\cline{1-5}
\colhead{} & \colhead{} & \colhead{} &
\colhead{initial} & \colhead{advanced} \\
\colhead{} & \colhead{Myr$^{-1}$} & \colhead{} &
\colhead{kyr$^{-1}$} & \colhead{yr$^{-1}$} 
}

\startdata
 1 & $23.2_{-18.5}^{+59.4}$     & 6.4 & $9.7_{-7.7}^{+24.9}$   & $52.2_{-41.6}^{+133.6}$\\
\\
 {\bf 6} &  $83.0_{-66.1}^{+209.1}$   & 6.4 & $34.8_{-27.7}^{+87.6}$  & $186.8_{-148.7}^{+470.5}$\\
\\
 9 & $7.9_{-6.3}^{+20.2}$      & 6.6 & $3.3_{-2.6}^{+8.4}$     & $17.7_{-14.1}^{+45.4}$\\
[0.07cm]
 10 & $23.3_{-18.4}^{+57.0}$    & 5.8  & $9.8_{-7.7}^{+23.9}$   & $52.4_{-41.3}^{+128.2}$\\
[0.07cm]
12 & $9.0_{-7.1}^{+21.9}$     & 6.0 & $3.8_{-3.0}^{+9.2}$    & $20.2_{-15.9}^{+49.4}$\\
[0.07cm]
 14 & $3.8_{-2.8}^{+9.4}$      & 5.4  & $1.6_{-1.2}^{+3.9}$    & $8.5_{-6.2}^{+21.1}$\\
\\
 15 & $223.7_{-180.6}^{+593.8}$ & 7.1 & $93.7_{-75.6}^{+248.6}$ & $503.2_{-406.3}^{+1336.0}$\\
[0.07cm]
 17 & $51.6_{-41.5}^{+135.3}$   & 6.9 & $21.6_{-17.4}^{+56.7}$   & $116.1_{-93.4}^{+304.4}$\\
[0.07cm]
 19 & $14.6_{-11.7}^{+38.2}$     & 7.0 & $6.1_{-4.9}^{+16.0}$    & $32.8_{-26.3}^{+86.0}$\\
[0.07cm]
\\
 20 & $89.0_{-70.8}^{+217.9}$  & 6.2 & $37.3_{-29.6}^{+91.2}$   & $200.3_{-159.3}^{+490.3}$\\
\enddata
\tablenotetext{a}{Model numbers correspond to KKL. Model 1 was used as a reference model in KKL. Model 6 is our reference model in this study (see text).}
\tablenotetext{b}{Increase rate factor compared to previous rates reported in KKL; IRF$\equiv {\cal R}_{\rm peak,new}/{\cal R}_{\rm peak,KKL}$.}
\end{deluxetable}
%%
%% (end of table)
%%%%%%%%%%%%%%%%%%%%%%%%%%%%%%%%%%%%%%%%%%%%%%%%%%%%%%%%%%%%%%%%%%%%%%%

We now explore our results for all other models considered in
KKL. Our main results are shown in Table 1, where we have included a subset of models that reflect the widest variations of the rates (as shown in KKL, variations in the space distribution of pulsars are not important). The
main conclusions that can be easily drawn are: (1) The increase factor
on the in-spiral rate is highly {\em robust} against all
systematic variations of the assumed pulsar models and is strongly
constrained in the range 5--7; this is consistent with 
but somewhat lower than the simple estimate presented in Burgay et al.\ (2003). (2) The shape of
the rate probability distribution also remains robust, but the
rate value at peak probability depends on the model assumptions in the
same way as described in detail in KKL (see Figs. 5--7 in KKL).

\section{PREDICTIONS FOR FUTURE DISCOVERIES}
 \label{sec:pmb} 
 
As already mentioned, long integration times combined with very short
binary orbital periods strongly select against the discovery
of new binary pulsars. Specifically, in the large-scale PMB survey (e.g. Manchester et al.\ 2001) with an integration time of 35\,min, the signal-to-noise ratio is severely reduced by Doppler 
smearing due to the pulsars' orbital motion. Acceleration searches in the
current re-analysis of the PMB survey (Faulkner et al.\ 2003) should 
significantly improve the detection efficiency to DNS binaries. 
 
Following Kalogera, Kim \& Lorimer (2003), we calculate the probability distribution that represents our expectation that the actual number of {\em DNS pulsars with merger times shorter than a Hubble time} ($N_{\rm obs}$)  that could be detected with the PMB survey takes on a particular value, given the current observations and assuming
that the reduction in flux due to Doppler smearing is corrected
perfectly. To illuminate the effect of the Doppler smearing we
calculate the average number of expected new discoveries akin to each
of the three known DNS binaries. 

We have shown before (Kalogera et al.\ 2003) that the probability distribution of the  expected observed number $N_{\rm obs}^i$ for each DNS pulsar sub-population $i$ (B1913+16,  B1534+12, and J0737--3039) is given by: 
 \begin{equation}
 P_i(N_{\rm obs})={\frac{{\beta_i}^{2}}{(1+\beta_i)^{2}}} { \frac{({N_{\rm
obs}}+1)}{(1+{\beta_i})^{N_{\rm obs}}} }, 
 \end{equation}
 where the constants $\beta_i$ are a measure of how less likely it is to detect pulsars without acceleration searches relative to with acceleration searches. For each sub-population, the mean values of $N_{\rm obs}$ can be calculated and we find them to be: 
   \begin{equation}
   \langle N_{\rm obs} \rangle_{\rm 1913} = 0.9, \hspace{0.3cm} 
   \langle N_{\rm obs} \rangle_{\rm 1534} = 1.2, \hspace{0.3cm}
   \langle N_{\rm obs} \rangle_{\rm 0737} = 1.9. 
   \end{equation}
 As expected, it is evident that the discovery of DNS pulsars in tight binaries like J0737--3039 would be most favored with acceleration searches. 
 
\begin{figure}
\begin{center}
\psfig{figure=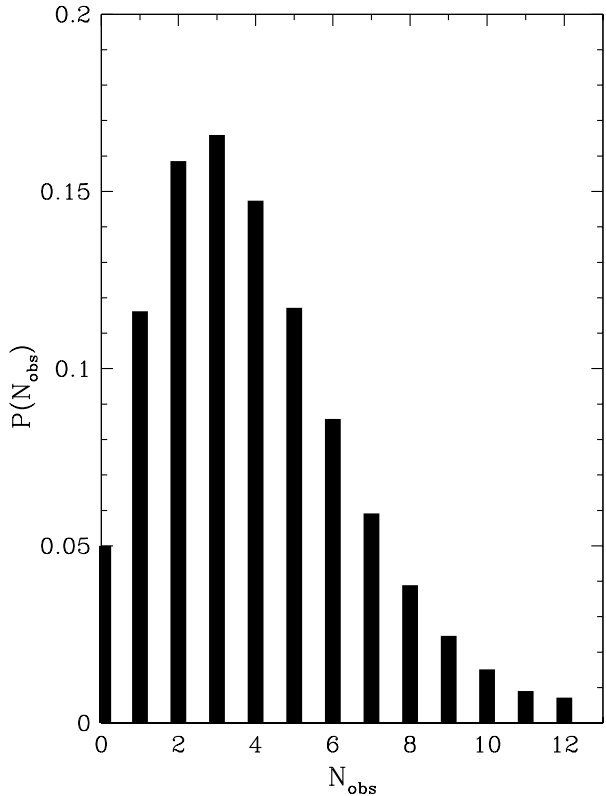,width=3.3in}
\caption{Probability density function of the predicted number of
observed DNS binary systems $N_{\rm obs}$ for the PMB survey, for
our reference model (model 6 in KKL).
The mean value is estimated to be $<$N$_{\rm obs}$>$=$4.0.
}
\label{fig:PMB}
\end{center}
\end{figure}

Following Kalogera et al.\ (2003), we can also calculate the
combined probability distribution of the expected number of DNS pulsars that can be detected with PMB acceleration searches in the future. The result is shown for our reference model in Figure \ref{fig:PMB}. The average combined number is 4.0 and the discovery of up to 2 (or 4) DNS systems has a probability equal to $\simeq 30$\% ($\simeq 63$\%). 
We conclude that, if the acceleration search can correct the Doppler smearing effect perfectly, then the PMB survey could be expected to detect an average of 3-4 DNS pulsars with pulse profile and orbital properties similar to any of the three already known systems.

\section{IN-SPIRAL EVENT RATES AND CONCLUSIONS}
\label{sec:discussion}

Estimates of DNS in-spiral rates have suffered from the small number of relativistic binaries known in our Galaxy, 
mainly because of the implicit assumption in all methods used so far that the observed sample represents 
the Galactic DNS population. Here we show that the recent discovery of the third 
relativistic binary in the Galactic field with binary properties (pulse-profile and orbital characteristics) significantly different from those of systems previously known reveals a new sub-population in the Galaxy. 
Consequently, it leads to a significant (by factors of 5--7) increase of the in-spiral rate estimates.

We now consider the implications of our revised rate estimates for
the detection of these events by LIGO and the other upcoming
gravitational wave interferometers. Since these instruments can detect
DNS in-spirals out to $\simeq 20$\,Mpc for initial LIGO 
($ \simeq 350$\,Mpc for advanced LIGO; Finn 2001), it is necessary
to extrapolate our Galactic event rate out to the Local Group. Using
the standard extrapolation of our reference model out to extragalactic
distances (Phinney 1991; KNST), we find the
most probable event rates for our reference model are one per 30 yr and one per 2 day, for
initial and advanced LIGO, respectively. At the 95\% confidence interval, the most optimistic predictions for the reference model are one event per 8\,yr and two events per day for initial and advanced LIGO, respectively. However, considering the full set of 27 models at 95\% confidence interval indicates that the respective rates can reach up to one event per 3 yr and five events per day, respectively. These results are quite encouraging, since, for initial LIGO in particular, this is the first time that DNS coalescence rate estimates are within an astrophysically relevant regime. Within a few years of LIGO operations, it should be possible to directly test these predictions and, in turn, place better constraints on the properties of binary radio pulsars and the cosmic population and evolution of DNS binaries. 

We also find that there is a significant probability (in excess of $\simeq$80\%) that when acceleration searches of the PMB survey are completed more than 2 binary pulsars could be detected. The increase of the observed sample is very important for the reduction of the uncertainties associated with the in-spiral rate estimates. We note, however, that the discovery of new systems that are {\em similar} to the three already known does not necessarily imply a significant increase in the rate estimates. Significant changes are expected in the case that new systems with pulse profiles or binary properties significantly different are discovered, as it is such systems that will reveal a new DNS sub-population in the Galaxy.

\acknowledgments
This work is partially supported by a David and Lucile Packard Science
\& Engineering Fellowship and a NSF Gravitational Physics grant
(0121420) to VK. DRL is a University Research fellow funded by the
Royal Society. He also acknowledges support by the Theoretical
Astrophysics Visitors' fund at Northwestern University. FC acknowledges support from NSF grant AST-02-05853 and a travel grant from NRAO.

\clearpage

NOTE FROM THE ERRATUM \\

In our original paper, we calculated the likely size of the Galactic
DNS population in two stages. First, we simulated the DNS distribution in the Galaxy.  
At this stage, in addition to storing the spatial properties and luminosities of the
model pulsars, we also computed their expected dispersion measures and
pulse scatter-broadening times using a model for the electron density
distribution. This information was stored to an intermediate file
for subsequent analysis by our simulation code in which the
simulated population was ``searched'' using detailed models
of the various pulsar surveys. Unfortunately, while the scatter-broadening times created 
in the first part of the calulation were saved in units of milliseconds, the survey
simulation code assumed them to be in seconds. 
This error led to an under-prediction of the number of pulsars in the
model observed samples, which in turn resulted in an overestimate of the true
number, and hence in-spiral rate, of DNS binaries in the Galaxy.

We have now repeated these calculations using the correct unit conversion and
find the Galactic DNS in-spiral rate ${\cal R}$ to be 83.0$_{-66.1}^{+209.1}$
Myr$^{-1}$ at a 95\% confidence interval (model 6), a factor of $\sim$2.2
smaller than in the original paper (see Fig. 1). The corresponding detection
rates for the initial and advanced LIGO detectors are: ${\cal R}_{\rm
det,ini}=(34.8_{-27.7}^{+87.6}) \times 10^{-3}$ yr$^{-1}$ and ${\cal
R}_{\rm det,adv}=186.8_{-148.7}^{+470.5}$ yr$^{-1}$, respectively.
For all models we consider, 
the estimated DNS in-spiral rates range between $4-224$ Myr$^{-1}$. 

However, we note that the increase rate factor (IRF) due to the discovery of
J0737--3039 remains essentially unaffected as the unit error cancels
out, since the IRF is a measure of {\it relative} changes in rate. 
The IRF is found to be in the range $5-7$ for all models under consideration. 
The revised results are summarized in Table 1.

This error also propagates through to our predictions for future
DNS discoveries in the Parkes multibeam survey.
Repeating the analysis in \S4 of our original paper, we find that the average
number of DNS with properties similar to those currently known
to be detected using full acceleration search processing is about four. 
The revised probability distribution of expected detections is shown in Fig.~2.

\end{document}